\documentclass[12pt,a4paper]{article}
\def \rstarbrack {\}_{\!{}_\star}}
\def \rodotbrack {]_{{}_\odot}}
\def \AO {{\widehat A}}
\def \BO {{\widehat B}}
\def \CO {{\widehat C}}
\def \HO {{\widehat H}}
\def \qO {{\hat q}}
\def \pO {{\hat p}}
\def \Wmap {{\cal W}}
\def \Winv {{\cal W}^{-1}}
\def \RO {{\widehat \rho}}
\def \roc {\rho_{\!{}_C}}
\def \roq {\rho_{\!{}_Q}}
\def \rohc {{\widehat \rho}_{\!{}_C}}
\def \rock {\rho_{\!{}_{CK}}}
\def \rohq {{\widehat \rho}_{\!{}_Q}}
\def \roqk {\rho_{\!{}_{QK}}}
\def \be {\begin{equation}}
\def \ee {\end{equation}}
\def \bea {\begin{eqnarray}}
\def \mea {\nonumber\\}
\def \eea {\end{eqnarray}}
\begin{document}
\title{Quantum Mechanics as an Approximation to 
Classical Mechanics in Hilbert Space}
\author{A.J. Bracken
\thanks{{\bf Email: ajb@maths.uq.edu.au}
On leave from Centre for Mathematical Physics, 
Department of Mathematics, University of
Queensland, Brisbane 4072, Australia. 
It is a pleasure to thank G. Cassinelli for his generous hospitality, 
and to thank him and J.G. Wood
for helpful discussions.  }
\\
DIFI, Universit\`a di Genova\\
Via Dodecaneso 33\\
Genova, 16146 ITALY
}
%\date{\today}
\maketitle
%\newpage
\begin{abstract}Classical mechanics is formulated 
in complex Hilbert space with the introduction
of a  commutative product of operators,  an 
antisymmetric bracket, and a quasidensity operator. 
These are analogues of the star product,
the Moyal bracket, and the Wigner function 
in the phase space formulation of quantum mechanics.  
Classical mechanics can now be viewed as a deformation of quantum mechanics. 
The forms of semiquantum approximations to classical mechanics
are indicated.   
\end{abstract}

%\newpage 

While our understanding of the relation between 
quantum mechanics and classical mechanics has 
steadily increased over the past 75 years, as 
a result of many studies from various points of
view (see \cite{berry}-\cite{ozorio} and references 
therein), few would claim that it is complete. 
Meanwhile, 
increasing attention has focussed on the
interface between the quantum and classical domains, 
because of
advances in experimental science and engineering, 
and the associated development of
`nanotechnology.' 

Classical mechanics is usually formulated in real, 
finite-dimensional phase space;
quantum mechanics in complex, infinite-dimensional Hilbert space.  
However, a completely equivalent reformulation of 
quantum mechanics in phase space is known \cite{weyl}-\cite{dubin},
which shows that 
quantum mechanics is a
deformation of classical mechanics \cite{bayen}, 
and which provides a natural setting for the 
formulation of semiclassical approximations 
\cite{wigner,berry,osborn}.
These allow us to explore
the interface between the two forms of mechanics when approached
from the
classical side.

In his remarkable 1946 paper,
Groenewold \cite{groenewold} indicated the alternative possibility of
reformulating classical mechanics as a quantum-like theory, 
with a quasidensity operator which is not positive-definite, 
although few details were given. This idea
has since been commented upon \cite{bayen} and explored in different ways \cite{jordan},
notably by Muga {\it et al.} \cite{muga}. 
It is not to be confused with the approach to
classical mechanics in the real 
Hilbert space of square-integrable phase space functions, initiated by
Koopman \cite{koopman}.
Here we show that, just as quantum
mechanics can be reformulated in phase space, 
so classical mechanics can be reformulated in
complex Hilbert space, in such a way that  
classical mechanics is seen as a
deformation of quantum mechanics.  
And now there arises the possibility of 
exploring the interface between quantum mechanics
and classical mechanics from the other side, the 
quantum side, with the systematic development of semiquantum approximations.

We limit discussion to a system with one linear degree of freedom.
All formulas below can
be generalized to many (possibly infinitely many!) 
degrees of freedom.  Our presentation is formal and heuristic; 
there is no attempt at mathematical
rigor.

A conservative classical 
system
is usually described in terms of  
functions (classical observables) $A_C(q,p)$ on
phase space, 
together with 
a probability density $\roc (q,p,t)$, 
characterizing the state of the system at time $t$,
with evolution equation 
\bea
\frac{\partial \roc }{\partial t}
&=&\{H_C,\roc \}_{P}\,
\equiv H_C\,J \,\roc \,,
\mea
J&=&
\frac{\partial ^L}{\partial q} 
\frac{\partial ^R}{\partial p} 
-
\frac{\partial ^L}{\partial p} 
\frac{\partial ^R}{\partial q} \,.
\label{liouville}
\eea
Here $H_C$ is the Hamiltonian function, 
$\{A,B\}_{P}$ denotes the Poisson bracket, and
the superscripts $L$ and $R$ indicate the directions in which the differential
operators act.
The expectation value of the classical observable $A_C(q,p)$ at time $t$ is
\be
\langle A_C\rangle (t)=
\int A_C(q,p)\roc  (q,p,t)\,dq dp\,.
\label{classexpec}
\ee
In (\ref{classexpec})
and below, integrals are over all real values of the variables of integration.

A conservative quantum system
is usually described 
in terms of  a complex Hilbert space
of 
square-integrable state functions $\psi (x)$.
Quantum observables are linear  
operators $\AO_Q$ acting on state functions as  
\be
(\AO_Q\psi)(x) = 
\int 
A_{QK}(x,y)\psi(y)\,dy\,,
\label{intopaction}
\ee
where $A_{QK}(x,y)$ is a complex-valued function, the kernel of $\AO_Q$.
In particular, the canonical coordinate and momentum operators $\qO$ and 
$\pO$ have  kernels  $x\delta(x-y)$ 
and $-i\hbar \delta'(x-y)$, respectively, where 
$\delta(x)$ is Dirac's `delta function.'
If the observable quantity is real, the corresponding operator is Hermitian\,:
$A_{QK}(x,y)=A_{QK}(y,x)^{*}$.   
An important example is the quantum density 
operator $\rohq (t)$, which has a kernel
\be
\roqk (x,y,t)=\sum_r p_r
\psi_r(x,t)\psi_r(y,t)^{*}
\label{rhokernel}
\ee
when the system is in a state described by 
the `mixture' of orthogonal state functions $\psi_r(x,t)$  with
associated probabilities $p_r$ at time $t$.
The quantum density operator is positive definite, with unit trace,
and the expectation value 
of the quantum observable $\AO_Q$ at time $t$ is  
\be
\langle \AO_Q\rangle (t) ={\rm Tr}(\AO_Q\rohq (t))\,.
\label{quantumexpec}
\ee
The evolution equation for $\rohq (t)$ is 
\be
\frac{\partial \rohq }{\partial t}=\frac{1}{i\hbar}[\HO_Q,\rohq ]\,,
\label{schrodinger}
\ee
where $\HO_Q$  is the Hamiltonian operator, 
and $[\AO_Q,\BO_Q]$ denotes the commutator.

In order to 
map the Hilbert space formulation of quantum mechanics into the
phase space formulation, 
the Weyl-Wigner transform $\Wmap$ is introduced.
For each quantum observable $\AO_Q$ with kernel $A_{QK}(x,y)$,
a corresponding
function $A_Q=\Wmap(\AO_Q)$ on phase space  is defined by setting
\be
A_Q(q,p)=
\int A_{QK}(q-x/2,q+x/2)\,e^{ipx/\hbar}\,dx\,.
\label{Adef}
\ee
If $\AO_Q$ is Hermitian, then $A_Q$ is real. 
The Wigner density function $\roq (t)=\Wmap(\rohq (t))
/(2\pi\hbar)$ is a particular case,
in terms of which the quantum expectation value 
(\ref{quantumexpec}) can be rewritten as
\be
\langle \AO_Q\rangle (t)=\int _{\Gamma} A_Q(q,p)\roq (q,p,t)\,dqdp\,.
\label{quantexpec2}
\ee
This has the appearance of the classical average (\ref{classexpec}), but while 
the Wigner function is real and normalised, it  is 
not in general nonnegative everywhere on phase space, and consequently
can  be interpreted only as a quasiprobability density.

In order to describe dynamics in the phase space formulation, the celebrated
star product 
and star (or Moyal) bracket 
of quantum phase
space functions   
are introduced
\cite{neumann,groenewold,moyal}\,:
\bea
A_Q\star B_Q&=&\Wmap(\AO_Q\BO_Q)\,,\quad\qquad\qquad
\mea
\{A_Q,B_Q\rstarbrack 
&=&\frac{1}{i\hbar}\, \Wmap([\AO_Q,\BO_Q])
\mea
&=&\frac{1}{i\hbar} (A_Q\star B_Q-B_Q\star A_Q)\,.
\label{stardef}
\eea
Then $q\star p= qp+i\hbar/2$, $p\star q=qp-i\hbar/2$, 
$q^2\star p^3= q^2p^3+3i\hbar q p^2 -3 \hbar^2 p$, {\it etc.}
The quantum evolution (\ref{schrodinger}) is now replaced by  
\be
\frac{\partial\roq (t)}{\partial t}=
\{H_Q,\roq \rstarbrack\,,\quad 
\label{quantevolution2}
\ee
where $H_Q=\Wmap(\HO_Q)$.

For suitably smooth $A_Q$ and $B_Q$, in particular polynomials in $q$ and $p$, 
it can be shown from (\ref{stardef}) that  
\be
\{A_Q,B_Q\rstarbrack =A_Q\,G\,B_Q\,,\quad G=\frac{2}{\hbar} 
\,\sin[\,\frac{\hbar}{2}J\,] \,,
\label{moyal1}
\ee
where 
the sine function is to interpreted by its 
Taylor series, and $J$ is as in (\ref{liouville}).
For more general $A_Q$, $B_Q$, such an expansion has only an aymptotic meaning, 
so that
(\ref{quantevolution2}) leads to 
\bea
\frac{\partial\roq (t)}{\partial t}
&\sim& H_Q\,J\,\roq -\frac{\hbar ^2}{3!2^2} 
 H_Q\,J ^3\,\roq
 \mea
 &+&\frac{\hbar ^4}{5!2^4} 
 H_Q\,J ^5\,\roq \dots\,,\quad
(\hbar\to 0)\,.
\label{quantevolution3}
\eea

Equations (\ref{quantexpec2}) and (\ref{quantevolution3}) are to be compared 
with their classical counterparts
(\ref{classexpec}) and (\ref{liouville}), which are `obtained' when $\hbar \to 0$.   
It is not our purpose here to discuss the 
subtle mathematical difficulties associated with
this limiting process \cite{berry,osborn}.
Suffice it to say that (\ref{quantexpec2}) and (\ref{quantevolution3}) 
form a natural starting point for  discussions of the classical limit, 
and of semiclassical approximations to quantum mechanics as $\hbar$ approaches $0$.

We now stand the foregoing on its head.
With each classical phase space function $A_C(q,p)$ we associate a linear operator 
$\AO_C=\Winv (A_C)$.  
This defines $\AO_C$ as the operator with kernel
\be
A_{CK}(x,y)= \frac{1}{2\pi\hbar}
\int
A_C([x+y]/2,p)\,e^{ip(x-y)/\hbar}\,dp\,.
\label{classkerneldef}
\ee
If $A_C$ is real, then $\AO_C$ is Hermitian.  
This is the usual Weyl mapping \cite{weyl} from functions to operators,
but our intention here is not to quantize, 
but to reformulate classical mechanics in complex Hilbert
space.  
It may then be objected that Planck's constant is not available to us in a 
classical theory.  We treat $\hbar$ for the moment as a parameter with dimensions of
action, whose value is to be specified at our convenience.

As a special case, we 
have the Groenewold density operator \cite{groenewold}
\be
\rohc (t)=2\pi\hbar 
\,\Winv(\roc (q,p,t))\,,
\label{rhoCdef}
\ee
This can be seen to be bounded, with unit trace,
but unlike a true quantum density operator, it is not always 
positive-definite.  Just as the Wigner function $\roq (q,p,t)$
is only a quasiprobability density,
so the Groenewold operator $\rohc (t)$ is only a quasidensity operator. 
But just as quantum averages can be calculated using the Wigner 
function  in the `classical' formula
(\ref{quantexpec2}),
so classical averages can be calculated using  $\rohc (t)$ 
in the `quantum' formula
\be
\langle A_C\rangle (t)={\rm Tr}(\AO_C\rohc (t))\,,
\label{classexpec2}
\ee
where $\AO_C$ is the operator corresponding to the classical function $A_C(q,p)$.  
These averages do not, of course, involve the parameter $\hbar$.

In order to describe classical dynamics in complex Hilbert space, 
we first introduce a distributive, associative and 
commutative `odot' product of operators,  
\be
\AO_C\odot\BO_C=\Winv (A_CB_C)=\BO_C\odot \AO_C\,.
\label{odotdef1}
\ee
Then for example, $\qO\odot \pO = \pO\odot 
\qO=(\qO\pO+\pO\qO)/2$, $\qO^2\odot \pO^3=
\pO^3\odot\qO^2= (\qO^2\pO^3 +2\qO\pO^3\qO +\pO^3\qO^2)/4$, {\it etc.}  
More generally, $\{\qO^k\pO^l\}\odot \{\qO^m\pO^n\}=
\{\qO^{k+m}\pO^{l+n}\}$, where $\{\qO^r\pO^s\}$ 
denotes the Weyl-ordered operator \cite{weyl,berezin}
corresponding
to the classical monomial $q^rp^s$.    
This follows from (\ref{odotdef1}) because $\{\qO^r\pO^s\}=\Winv (q^rp^s)$.   

Most generally, it can be  
seen from (\ref{classkerneldef}) 
that the kernels of the operators $\AO_C$, 
$\BO_C$ and $\AO_C\odot \BO_C$ are related by 
\bea
&&\qquad\qquad(\AO_C\odot\BO_C)_K(x,y)=
(\BO_C\odot\AO_C)_K(x,y)
\mea
&&=\int
A_{CK}\big([3x+y-2u]/4,[x+3y+2u]/4\big)
\mea
&&\,\,\times B_{CK}
\big([3x+y+2u]/4,[x+3y-2u]/4\big)\,du\,.
\label{odotprod2}
\eea

It is helpful to introduce the notations
\bea
A_q=\partial A/\partial q\,,\quad A_{qp}=\partial ^2 A/\partial q \partial p\,,
\dots
\mea
\AO_q=\frac{1}{i\hbar}[\AO,\pO]\,,\quad \AO_{qp}=
\big(\frac{1}{i\hbar}\big)^2[\qO,[\AO,\pO]]\,,\dots
\label{notation}
\eea
and to note that,  because $A_{qp}=q\,G\,(A\,G\,p)$, {\it etc.}, and 
\be
\Winv(A\,G\,B)=\frac{1}{i\hbar}[\AO,\BO]\,,
\label{Gmap}
\ee
we have $\Winv(A_{qp})=\AO_{qp}$, {\it etc.}
In (\ref{notation}), $\qO$ and $\pO$ are the usual canonical operators, except 
with commutator involving the parameter
$\hbar$, whose value has not yet been fixed.   

To describe classical dynamics, we need to introduce 
a new bracket, equal except for a 
convenient factor to the image of the Poisson
bracket under the inverse
Weyl-Wigner transform. 
We set 
\bea
[\AO_C,\BO_C\rodotbrack&=&i\hbar\, \Winv (\{A_C,B_C\}_{P})
\mea
&=&
i\hbar\big(\AO_{Cq}\odot \BO_{Cp}-\AO_{Cp}\odot \BO_{Cq}\big)\,.
\label{newbrackdef}
\eea

Now the classical evolution equation (\ref{liouville}) is replaced by 
\be
\frac{\partial \rohc }{\partial t}=\frac{1}{i\hbar}[ \HO_C,\rohc \rodotbrack\,.
\label{classevolution2}
\ee
We emphasize that this reformulation of classical mechanics in
terms of linear operators on 
Hilbert space, incorporating the arbitrary parameter $\hbar$, and
with key equations (\ref{classexpec2}) and (\ref{classevolution2}), 
is entirely equivalent to the usual phase space formulation.   
We can go back and forth between the two descriptions with the 
help of the Weyl-Wigner transform $\Wmap$ 
and its inverse
$\Winv$.

Next we make an
expansion of the odot bracket,  analogous to the expansion (\ref{moyal1}).
Noting that
\be
\theta = \sin\theta (1+\theta^2/6 + 7\theta^4/360-\dots\,,\quad |\theta|<\pi\,,
\label{thetaexpand}
\ee
we write 
\bea
&&\qquad A\,J\,B =
A\, G\,B+ \frac{1}{6}\left(\frac{\hbar}{2}\right)^2 A\,J^2\,G\,B
\mea
&&\quad+\frac{7}{360}\left(\frac{\hbar}{2}\right)^4 A\,J^4\,G\,B -\dots 
=A\,G\,B 
\mea
&&+
\frac{1}{6}\left(\frac{\hbar}{2}\right)^2\big(A_{qq}\,G\,B_{pp}
-2 A_{qp}\,G\,B_{qp}
+A_{pp}\,G\,B_{qq}\big)+\dots
\mea
\label{pbgm}
\eea
and then,  applying $\Winv$ to both sides, 
\bea
&&\qquad\qquad[\AO,\BO\rodotbrack
=
[\AO,\BO]
\mea
&&+
\frac{1}{6}\left(\frac{\hbar}{2}\right)^2
\big(
[\AO_{qq},\BO_{pp}] -2[\AO_{qp},\BO_{qp}]+[\AO_{pp},\BO_{qq}]\big)+\dots
\mea
\label{newbrack2}
\eea
The series (\ref{pbgm}) and (\ref{newbrack2}) 
terminate if at least one of $A$ and $B$ is a
polynomial in $q$ and
$p$.  For more general $A$ and $B$, 
we may expect that the  series 
have well-defined meanings as asymptotic expansions when $\hbar\to 0$.

The classical evolution (\ref{liouville}) then  takes the form 
\bea
&&\frac{\partial \rohc }{\partial t}
\sim
\frac{1}{i\hbar}[\HO_C,\rohc ]
\mea
&&-\frac{i\hbar}{24}
\big(
[\HO_{Cqq},\RO_{\!{}_{Cpp}}]
-2
[\HO_{Cqp},\RO_{\!{}_{Cqp}}]
+
[\HO_{Cpp},\RO_{\!{}_{Cqq}}]\big)
\mea
&&-\dots\,, \quad (\hbar\to 0)\,. 
\label{classevolution3}
\eea
If $H_C$ is a polynomial in $q$ and $p$, then this series 
terminates and the
asymptotic result becomes exact.

If $H_C=H(q,p)=p^2/(2m)+V(q)$, then (\ref{classevolution3}) reduces to
\bea
&&\frac{\partial \rohc }{\partial t}
\sim
\frac{1}{i\hbar}[H(\qO,\pO),\rohc ]
-\frac{i\hbar}{24}[V''(\qO),\RO_{\!{}_{Cpp}}]
\mea
&&-\frac{7i\hbar ^3}{5760}[V^{(iv)}(\qO),\RO_{\!{}_{Cpppp}}]
+\dots\,,\quad (\hbar\to 0)\,, 
\label{nonrelcase}
\eea
which is an analogue of Wigner's equation for 
the evolution of his density function \cite{wigner}.

If we {\it now} identify $\hbar$ with Planck's constant, 
we see that the equations (\ref{quantumexpec}) 
and (\ref{schrodinger}) of quantum mechanics  
are  obtained formally as $\hbar$ approaches $0$, and 
that classical mechanics can be regarded as a deformation of quantum mechanics,
with deformation parameter $\hbar$.  Most interesting is that 
(\ref{classexpec2}), taken with (\ref{classevolution3}) or (\ref{nonrelcase}), may be
expected to form a suitable starting point for semiquantum approximations to classical
mechanics, analogous to semiclassical approximations to quantum mechanics.  
Successive approximations will be associated with successively later terminations of the series
(\ref{classevolution3}) or (\ref{nonrelcase}).  Note that $\hbar$ may then appear in the corresponding
approximations to
classical averages calculated using (\ref{classexpec2}), in contrast to its non-appearance in 
exact classical averages.

These results may seem paradoxical.  We have introduced 
$\hbar$ into a reformulation of
classical mechanics, without affecting its predictions 
in any way, and see that as that parameter
approaches $0$, the equations of quantum mechanics emerge.  
Usually we say, speaking loosely, that
classical mechanics is obtained from quantum mechanics as $\hbar$ approaches $0$. 
Viewing things from the perspective provided by 
the above results, we argue that it is more
appropriate to say 
that classical mechanics and quantum mechanics become
asymptotically equivalent as $\hbar\to 0$\,: 
the interface can be approached from either side.

We close with  a few remarks about interesting issues arising:

$\bullet$
The fundamental importance in quantum mechanics 
of the spectra of selfadjoint operators, the superposability of state
functions, and the nonunitary change in the density 
operator following a measurement, are obscured
in the phase space formulation.  They underlie the 
determination of averages (\ref{quantexpec2}) and of
initial values of Wigner functions.  
On the other hand, the complex Hilbert space formulation of 
classical mechanics begs the question: What are 
the relevance to classical mechanics, when formulated in this way, 
of operator spectra and the superposability of complex vectors?

$\bullet$
Consider a normalized classical density at some fixed time given by 
\be
\roc (q,p)=\frac{\sqrt{\alpha\beta}}{\pi}e^{-(\alpha   q^2 +\beta  p^2)}\,.
\label{classdensity}
\ee
The corresponding 
quasidensity operator $\rohc $  has  kernel  
\be
\rock (x,y)=
\sqrt{\frac{\alpha}{\pi}}e^{-\alpha(x+y)^2/4} e^{-(x-y)^2/(4\beta\hbar^2)}\,.
\label{kernel1}
\ee
It is easy to check that this operator is bounded, with  unit trace, 
but it is not in general positive definite.
If $\alpha\beta=1/(\hbar)^2$, so that the product of the uncertainties  
of $q$ and $p$ equals $\hbar/2$,
the kernel  factorizes:  
\be
\rock =\psi(x)\psi(y)^{*}\,,\quad \psi(x)=
\sqrt[4]{\frac{\alpha}{\pi}}\,e^{-\alpha x^2/2}\,,
\label{kernel2}
\ee
and the operator has the form of 
a true, positive-definite density operator, corresponding to the pure coherent
state
$\psi (x)$.  More generally, a little thought shows that
the only positive-definite quasidensity operators 
are those corresponding to convex linear 
combinations of
Gaussian $\roc (q,p)$, each with the product of the uncertainties  
in $q$ and $p$ equal to $\hbar/2$.
At the other extreme, as $\alpha\to\infty$ and $\beta\to\infty$, then
$\roc (q,p)\to\delta(q)\delta(p)$ and $\rock (x,y)\to2
\delta(x+y)$.     This defines the
starting point of a classical trajectory, 
as described in the Hilbert space formulation.   

$\bullet$
Consider a classical system exhibiting chaos \cite{gutzwiller},
for example the Henon-Heiles oscillator 
with 2 degrees of freedom and Hamiltonian
\bea
&&\qquad\qquad H_C=H(q_1,q_2,p_1,p_2)
\mea
&&=a(p_1{}^2+p_2{}^2)+b(q_1{}^2+q_2{}^2)+cq_1(3q_2{}^2-q_1{}^2)
\,.
\label{henon}
\eea
This system is described in Hilbert space by (\ref{classexpec2}) and 
(the obvious generalization of)
(\ref{nonrelcase}), with the series terminating 
after the terms of order $\hbar$.
If we choose a Gaussian initial density, generalizing 
(\ref{classdensity}), with arbitrarily small
uncertainties in the dynamical variables then, 
with the help of a computer, we can in principle
track  the average evolution
of the classical system, again with arbitrarily small uncertainties, 
and even if the motion is chaotic,
while working in the Hilbert space formalism.   
This  is remarkable because in the leading `quantum approximation,' 
obtained by neglecting the terms of order
$\hbar$ in (\ref{nonrelcase}), the classical chaos 
is suppressed \cite{berry,gutzwiller}.

$\bullet$
The new bracket has the `odot
derivation property' and `odot Jacobi identity,' 
which it inherits from the Poisson bracket:  
\bea
[\AO,\BO\odot \CO\rodotbrack = \CO \odot 
[\AO,\BO\rodotbrack +\BO\odot [\AO,\CO\rodotbrack
\mea
\,[[\AO,\BO\rodotbrack,\CO\rodotbrack
+[[\BO,\CO\rodotbrack,\AO\rodotbrack
+[[\CO,\AO\rodotbrack,\BO\rodotbrack=0\,.
\label{jacobi}
\eea
Poisson algebras of phase space functions, and associated groups, 
should translate into interesting odot
operator structures 
in Hilbert space.


\begin{thebibliography}{99}


\bibitem{berry}
M.V. Berry, 
Philos. Trans. R. Soc. Lond. A {\bf 287}, 237--271 (1977); 
Proc. R. Soc. Lond. A {\bf 413}, 183--198
(1987). 

\bibitem{fedoriuk}
V.P. Maslov and M.V.   Fedoriuk,  
{\em Semiclassical Approximation in Quantum Mechanics},
(Reidel, Dordrecht, 1981).


\bibitem{gutzwiller}
M.C. Gutzwiller, {\em Chaos in Classical and Quantum Mechanics}, 
(Springer-Verlag, New York,
1990). 

\bibitem{osborn}
T.A. Osborn and F.H. Molzahn, 
Ann. Phys. (NY) {\bf 241}, 79--127 (1995). 

\bibitem{interface}
N. Fr\"oman and P.O. Fr\"oman, {\em Phase Integral Method}, 
(Springer-Verlag, New York, 1996).

\bibitem{percival}
I. Percival, {\em Quantum State Diffusion}, 
(Cambridge University Press, 1998). 

\bibitem{ozorio}
A.M. Ozorio de Almeida, Phys. Rep. {\bf 295}, 265--342 (1998). 

\bibitem
{weyl}
H. Weyl,
Zeitschr. Phys. {\bf 46}, 1--46 (1927); {\em The Theory of
Groups and Quantum Mechanics}, (Dover, New York, 1931), p. 274.

\bibitem{neumann}
J. von Neumann,  Math. Ann. {\bf 104}, 570--578 (1931).  

\bibitem{wigner}
E.P. Wigner, 
Phys. Rev. {\bf 40}, 749--759 (1932).

\bibitem
{groenewold}
H. Groenewold, 
Physica {\bf 12}, 405--460 (1946).

\bibitem
{moyal}
J.E. Moyal,
Proc. Camb. Phil. Soc. {\bf 45}, 99--124 (1949).

\bibitem{berezin}
F.A. Berezin and M.A. \v Subin, in B.Sz. Nagy (Ed.), 
{\em Hilbert Space Operators and Operator
Algebras.  Colloquia Mathematicae Societatis Janos Bolyai 5.}
(North Holland, Amsterdam, 1972), pp. 21--52. 

\bibitem
{bayen}
F. Bayen, M. Flato,  C. Fronsdal, A. Lichnerowicz and D. Sternheimer, 
Ann. Phys. (N.Y.) {\bf 111},
61-110; 111--151 (1978).

\bibitem{dubin}
D.A. Dubin,  M.A. Hennings  and T.B. Smith,  
{\em Mathematical Aspects of Weyl Quantization and
Phase}, (World Scientific, Singapore, 2000).

\bibitem{jordan}
T.F. Jordan and E.C.G. Sudarshan,  Rev. Mod. Phys. {\bf 33}, 515--524 (1961). 

\bibitem{muga}
J.G. Muga and R.F. Snider, Europhys. Lett. {\bf 19}, 569--573 (1992);
R. Sala and J.G. Muga, Phys. Lett. {\bf A192}, 180--184 (1994).

\bibitem{koopman}
B.O. Koopman, Proc. Natl. Acad. Sc. USA {\bf 17}, 315--318 (1931). 

\end{thebibliography}
\end{document}